\documentclass[preprint,12pt]{elsarticle}
\usepackage{graphics}
\usepackage{mathrsfs}
\usepackage{amsmath}

\def\Vec#1{\mbox{\boldmath $#1$}}
\def\+{\mbox{\boldmath $+$}}
\def\-{\mbox{\boldmath $-$}}
\journal{Physica A}

\begin{document}

\begin{frontmatter}

\title{Understanding quantum entanglement \\ by thermo field dynamics}


\author[1]{Yoichiro Hashizume}
\address[1]{Tokyo University of Science,\\ 6-3-1 Niijuku, Katsushika-ku, Tokyo, 125-8585, Japan}
\ead{hashizume@rs.tus.ac.jp}
\author[2]{Masuo Suzuki}
\address[2]{Computational Astrophysics Laboratory, RIKEN,\\ 2-1 Hirosawa, Wako, Saitama, 351-0198, Japan}
\ead{masuo.suzuki@riken.jp}

\begin{abstract}
We propose a new method to understand quantum entanglement using the thermo field dynamics (TFD) described by a double Hilbert space.
The entanglement states show a quantum-mechanically complicated behavior.
Our new method using TFD makes it easy to understand the entanglement states, because the states in the tilde space in TFD play a role of tracer of the initial states.
For our new treatment, we define an extended density matrix on the double Hilbert space.
From this study, we make a general formulation of this extended density matrix and examine some simple cases using this formulation.
Consequently, we have found that we can distinguish intrinsic quantum entanglement from the thermal fluctuations included in the definition of the ordinary quantum entanglement at finite temperatures.
Through the above examination, our method using TFD can be applied not only to equilibrium states but also to non-equilibrium states.
This is shown using some simple finite systems in the present paper.
\end{abstract}


\end{frontmatter}

\section{Introduction}
\label{intro}
We propose a new method to understand quantum entanglement using the thermo field dynamics (TFD).
The TFD is very convenient to understand entanglement states, because it focuses on the state directly.
In this section, we give a simple introduction of TFD as well as quantum entanglement.

\subsection{Thermo Field Dynamics}
The thermal average $\langle A\rangle_{\text{eq}}$ of a physical quantity $A$ is expressed by $\langle A\rangle_{\text{eq}}={\text{Tr}}A\rho(\beta)$, using $\rho(\beta)=e^{-\beta\mathcal{H}}/Z(\beta)$, where $Z(\beta)$ denotes the partition function.
On the other hand, the quantum expectation value $\langle A\rangle$ is expressed by $\langle A\rangle=\langle \phi|A|\phi\rangle$ using the state vector $|\phi\rangle$.
In TFD, the above two expressions are combined on the basis of the extended concepts of states [1-6].
While the ordinary states are expressed as a state vector defined in a Hilbert space, TFD requires a ``statistical'' state vector defined in the double Hilbert space which is defined as a direct product of the original space and its isomorphic space (namely, tilde space).
Here, when we choose a set of bases $\{|n\rangle\}$ in the Hilbert space, the bases of the tilde space are expressed as $\{|\tilde{n}\rangle \}$ [1-6].
Then, the bases of the double Hilbert space are shown as $\{ |n\rangle \otimes|\tilde{m}\rangle \}(\equiv\{|n\rangle|\tilde{m}\rangle\}\text{ or } \{|n,\tilde{m}\rangle\})$ [1-6].
The statistical states were originally (Fano\cite{1}, Prigogine\cite{2}, Takahashi-Umezawa\cite{3}) defined as
\begin{equation}
|\Psi(\beta)\rangle=\frac{1}{\sqrt{Z(\beta)}}e^{-\beta\mathcal{H}/2}|I\rangle;\quad |I\rangle\equiv\sum_{n}|n\rangle|\tilde{n}\rangle\equiv\sum_{n}|n,\tilde{n}\rangle\label{eq1}
\end{equation}
in the double Hilbert space, using the eigenstates $\{|n\rangle\}$ of the Hamiltonian $\mathcal{H}$, namely $\mathcal{H}|n\rangle=E_n|n\rangle$.
Then, the average $\langle A\rangle_{\text{eq}}$ of the physical quantity $A$ is expressed\cite{4,5} by the expectation value $\langle \Psi(\beta)|A|\Psi(\beta)\rangle$ in TFD as
\begin{align}
\langle \Psi(\beta)| A |\Psi(\beta)\rangle&=\sum_{n}\sum_{m}\frac{1}{Z(\beta)}\langle n|e^{-\beta\mathcal{H}/2}A e^{-\beta\mathcal{H}/2} |m\rangle\langle \tilde{n}|\tilde{m} \rangle\notag
\\
&=\sum_{n}\sum_{m}\frac{1}{Z(\beta)}\langle n|e^{-\beta\mathcal{H}/2}A e^{-\beta\mathcal{H}/2} |m\rangle\delta_{n,m}\notag
\\
&=\sum_n\frac{1}{Z(\beta)}e^{-\beta E_n}\langle n|A|n\rangle =\langle A\rangle _{\text{eq}}.\label{eq2}
\end{align}
While the derivation of (\ref{eq2}) in the original papers used a set of eigenstates $\{|n\rangle \}$ for the original definition of $|I\rangle$, one of the authors (M.S.) showed that the state $|I\rangle $ is invariant for any orthogonal complete set $\{|\alpha\rangle \}$\cite{4,5}.
For the state $|I\rangle=\sum_{n}|n\rangle|\tilde{n}\rangle$ expressed by the eigenstates $\{|n\rangle \}$, the unitary transformation
\begin{equation}
|n\rangle=\sum_{\alpha}U_{n,\alpha}|\alpha\rangle \text{ and } |\tilde{n}\rangle=\sum_{\alpha}U^{*}_{n,\alpha}|\tilde{\alpha}\rangle\label{eq3}
\end{equation}
gives the following transformation of $|I\rangle$:
\begin{align}
|I\rangle &=\sum_{n}|n\rangle|\tilde{n}\rangle
=\sum_{n}\sum_{\alpha}\sum_{\alpha'}U_{n\alpha}U^{*}_{n\alpha'}|\alpha\rangle|\tilde{\alpha'}\rangle \notag
\\
&=\sum_{\alpha}\sum_{\alpha'}\left(\sum_{n}U_{n\alpha}U^{*}_{n\alpha'}\right)|\alpha\rangle|\tilde{\alpha'}\rangle
=\sum_{\alpha}\sum_{\alpha'}\delta_{\alpha,\alpha'}|\alpha\rangle|\tilde{\alpha'}\rangle
=\sum_{\alpha}|\alpha\rangle|\tilde{\alpha}\rangle. \label{eq4}
\end{align}
Then, the statistical state vector $|\Psi\rangle$ does not depend on any representation\cite{4,5}.
This is called ``the general representation theorem" of TFD.
This means that not only the average but also the state itself do not depend on the basis $\{|\alpha\rangle\}$.
The above theorem is very important, because it makes it possible to study any state using TFD even in non-equilibrium systems.

In addition, the time evolution of the statistical state $|\Psi(t)\rangle $ is described by the following differential equation [4-6];
\begin{equation}
i\hbar\frac{\partial}{\partial t}|\Psi(t)\rangle=\hat{\mathcal{H}}|\Psi(t)\rangle,\quad \hat{\mathcal{H}}=\mathcal{H}(t)-\tilde{\mathcal{H}}(t).\label{eq5}
\end{equation}
Here, the tilde Hamiltonian $\tilde{\mathcal{H}}$ is an operator defined in the tilde space.
Consequently, it operates only to the tilde space elements.

As shown in the above discussion, the TFD formulation makes it possible to treat quantum {\it states} directly and this formulation is very useful for analyzing thermal quantum states.
In fact, it was applied to clarifying the existence of resonating valence bond (RVB) states in anti-ferromagnetic triangular lattice models \cite{6}, to preforming the density matrix renormalization group method (DMRG) for quantum systems including frustration \cite{7} and to analyzing the state of black holes \cite{8}.

\subsection{Quantum Entanglement}
Several states are entangled essentially through non-separable quantum fluctuations.
A typical example of it is seen in the singlet state.
These behaviors of quantum states are called ``quantum entanglements''.
Traditionally, the entanglement entropy is used as a measure of the strength of quantum entanglements.
The quantum entanglement plays an important role in quantum computations \cite{9,10}, and it is useful in applications of the AdS/CFT correspondence [12-16].
On the other hand, in some statistical studies of quantum spin systems, the entanglement entropy is used as an order parameter [17-23].
In addition, how the entanglement entropy corresponds to the classical one has been studied \cite{24} using the Suzuki-Trotter transformation \cite{25}.

To define the entanglement entropy, the relevant system is divided into two partial systems A and B.
Then the entanglement entropy $S_{\text{A}}$ is defined as
\begin{equation}
S_{\text{A}}=-k_{\text{B}}{\text{Tr}}_{\text{A}} \rho_{\text{A}}\log\rho_{\text{A}}{\text{ with }}\ \rho_{\text{A}}={\text{Tr}}_{\text{B}}\rho_{\text{A+B}},\label{eq6}
\end{equation}
where $\rho_{\text{A+B}},\rho_{\text{A}}$ and $\rho_{\text{B}}$ denote the density matrices of the total system, the partial system A and the partial system B, respectively.
Here, ${\text{Tr}_{\text{A}}}$ and ${\text{Tr}_{\text{B}}}$ correspond to the variables of the systems A and B, respectively.
From the definition of the entanglement entropy (\ref{eq6}), it is easily understood that $S_{\text{A}}$ includes the original fluctuation of the partial system A.
The above studies [9-24] do not separate the fluctuations, namely the original fluctuations and the entanglement fluctuations.
In addition, the entanglement entropy is not a physical parameter defined by eigenvalues of unitary operators, such as magnetization, but a status of states.
Then the TFD is well applied to the study of this entropy.

In the present paper, we introduce in section 2 an extended density matrix of the TFD state, and examined in sections 3 the entanglement entropy for some typical cases including non-equilibrium systems.
In section 4, we give the summary and discussions.

\section{Extended density matrix in double Hilbert space}
The extended density matrix $\hat{\rho}$ is defined in the double Hilbert space as follows:
\begin{equation}
\hat{\rho}\equiv |\Psi\rangle\langle \Psi|, \quad |\Psi\rangle=\rho^{1/2}|I\rangle,\label{eq7}
\end{equation}
using the ordinary density matrix $\rho$ in a Hilbert space.
Here, $\rho^{1/2}$ satisfies the condition $\rho=(\rho^{1/2})^2$.
The state $|I\rangle$ is denoted as $|I\rangle=\sum_{\alpha}|\alpha,\tilde{\alpha}\rangle=\sum_{\alpha}|\alpha\rangle |\tilde{\alpha}\rangle$ for any orthogonal complete set $\{|\alpha\rangle\}$ in the Hilbert space \cite{4,5}.
For example, the statistical state vector $|\Psi\rangle$ in thermal equilibrium states is expressed as Eq.(\ref{eq1}).
Then, the extended density matrix $\hat{\rho}(\beta)$ is expressed as
\begin{align}
\hat{\rho}(\beta)&=\frac{1}{Z(\beta)}\left(e^{-\beta\mathcal{H}/2}\sum_{\alpha}|\alpha,\tilde{\alpha} \rangle \right)\left( \sum_{\alpha'}\langle\alpha',\tilde{\alpha}'| e^{-\beta\mathcal{H}/2} \right)\notag
\\
&=\frac{1}{Z(\beta)}\sum_{\alpha,\alpha'} \left( e^{-\beta\mathcal{H}/2}|\alpha\rangle\langle\alpha'|  e^{-\beta\mathcal{H}/2} \right) |\tilde{\alpha}\rangle\langle\tilde{\alpha}'|.\label{eq8}
\end{align}
When we treat the non-equilibrium systems, the density matrix $\rho$ includes the time dependence, namely $\rho=\rho(t)$.

The extended density matrix $\hat{\rho}$ satisfies the following plausible conditions.
\begin{enumerate}
\item[i)] Let us take the trace of variables in the tilde space as follows:
\begin{align}
\tilde{\text{Tr}}\hat{\rho} &\equiv \sum_{l}\langle \tilde{l}|\hat{\rho}|\tilde{l}\rangle
=
\sum_{l}\sum_{n,m} \rho^{1/2} |n\rangle\langle m|\left(\rho^{1/2}\right)^{\dagger}\langle \tilde{l}|\tilde{n}\rangle\langle \tilde{m}|\tilde{l}\rangle\notag
\\
&=\sum_{l}\sum_{n,m} \rho^{1/2} |n\rangle\langle m|\left(\rho^{1/2}\right)\delta_{l,n}\delta_{m,l}\notag
\\
&=\sum_{l}\rho^{1/2}|l\rangle\langle l|\rho^{1/2}\notag
\\
&=\rho^{1/2}\left( \sum_{l}|l\rangle\langle l | \right)\rho^{1/2}=\rho^{1/2}\Vec{1}\rho^{1/2}=\rho.\label{eq9}
\end{align}
Thus we obtain the ordinary density matrix $\rho$ by taking the trace of variables in the tilde space.

\item[ii)] The statistical state $|\Psi(t)\rangle$ satisfies the time evolution equation as shown in Eq.(5).
Then, the time differentiation of $\hat{\rho}(t)$ yields
\begin{align}
\frac{\partial}{\partial t}\hat{\rho}(t)&=\frac{\partial}{\partial t}\left(|\Psi(t)\rangle\langle\Psi(t)| \right)\notag
\\
&=|\Psi(t)\rangle\left(\frac{\partial}{\partial t}\langle\Psi(t)|\right)+\left(\frac{\partial}{\partial t}|\Psi(t)\rangle\right)\langle\Psi(t)|\notag
\\
&=\frac{1}{i\hbar}[\hat{\mathcal{H}}(t),\hat{\rho}(t)].\label{eq10}
\end{align}
That is, the von-Neumann equation holds even in the double Hilbert space using the extended operators $\hat{\rho}$ and $\hat{\mathcal{H}}$.
\end{enumerate}

\begin{figure}[hbpt]
\begin{center}
\includegraphics[width=7cm]{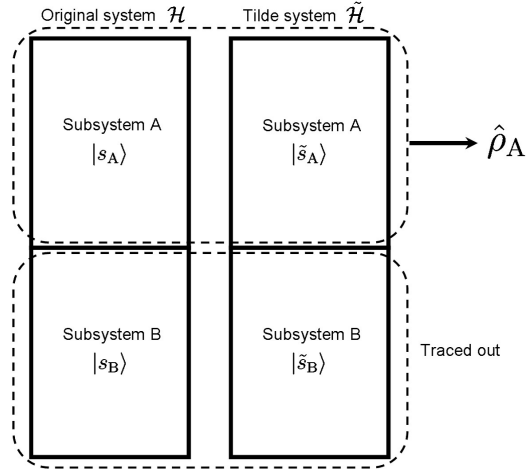}
\end{center}
\caption{The schematic model to obtain the density matrix $\hat{\rho}_{\text{A}}$. It is important to distinguish the states $|s_{\text{A}}\rangle,|s_{\text{B}}\rangle,|\tilde{s}_{\text{A}}\rangle$ and $|\tilde{s}_{\text{B}}\rangle$. There does not exist any interaction between the two systems $\mathcal{H}$ and $\tilde{\mathcal{H}}$. The parameters of the subsystem B in the original and tilde systems are traced out to obtain the density matrix $\hat{\rho}_{\text{A}}$.}
\label{fig0}
\end{figure}

Using the above extended density matrix $\hat{\rho}$, we make a general formulation to understand the entanglement states.
First, the state of the total system $|s\rangle$ is denoted by the direct product $|s\rangle=|s_{\text{A}},s_{\text{B}}\rangle=|s_{\text{A}}\rangle|s_{\text{B}}\rangle$, where $|s_{\text{A}}\rangle$ and $|s_{\text{B}}\rangle$ denote the states of subsystems A and B, respectively.
Using the general representation theorem [4,5], the statistical state can be expressed as
\begin{align}
|\Psi\rangle&=\sum_{s}\rho^{1/2}|s,\tilde{s}\rangle\notag
\\
&=\sum_{s_{\text{A}},s_{\text{B}}}\rho^{1/2}|s_{\text{A}},s_{\text{B}}\rangle |\tilde{s}_{\text{A}},\tilde{s}_{\text{B}}\rangle=\sum_{s_{\text{A}},s_{\text{B}}}\rho^{1/2}|s_{\text{A}},\tilde{s}_{\text{A}}\rangle |s_{\text{B}},\tilde{s}_{\text{B}}\rangle,\label{eq11}
\end{align}
with the states $\{|s\rangle\}$.
Thus, the renormalized extended density matrix $\hat{\rho}_{\text{A}}$ is expressed by tracing on the variables in the subsystem B as follows:
\begin{equation}
\hat{\rho}_{\text{A}}\equiv {\text{Tr}}_{\text{B}}\hat{\rho}\equiv \sum_{\gamma_{\text{B}},\tilde{\gamma}'_{\text{B}}}\langle \gamma_{\text{B}},\tilde{\gamma}'_{\text{B}}|\hat{\rho}|\gamma_{\text{B}},\tilde{\gamma}'_{\text{B}}\rangle. \label{eq12}
\end{equation}
The schematic model to obtain the above density matrix $\hat{\rho}_{\text{A}}$ is shown in Fig.{\ref{fig0}}.

Furthermore, the ordinary density matrix $\rho^{1/2}$ is expressed using the matrix elements $\{a_{\alpha_{\text{A}},\beta_{\text{B}},\alpha'_{\text{A}},\beta'_{\text{B}}}\}$ as
\begin{equation}
\rho^{1/2}\equiv \sum_{\alpha_{\text{A}},\beta_{\text{B}},\alpha'_{\text{A}},\beta'_{\text{B}}}a_{\alpha_{\text{A}},\beta_{\text{B}},\alpha'_{\text{A}},\beta'_{\text{B}}}|\alpha_{\text{A}},\beta_{\text{B}}\rangle\langle \alpha'_{\text{A}},\beta'_{\text{B}}|.\label{eq13}
\end{equation}
Then, by inserting Eqs.(\ref{eq11}) and (\ref{eq13}) into Eq.(\ref{eq12}), we obtain
\begin{align}
\hat{\rho}_{\text{A}}&=\sum_{\alpha_{\text{A}},\beta_{\text{A}},\alpha'_{\text{A}},\beta'_{\text{A}}}
b_{\alpha_{\text{A}},\beta_{\text{A}},\alpha'_{\text{A}},\beta'_{\text{A}}}
|\alpha_{\text{A}},\tilde{\beta}_{\text{A}}\rangle\langle\alpha'_{\text{A}},\tilde{\beta}'_{\text{A}}|\notag
\\
&=\sum_{\alpha_{\text{A}},\beta_{\text{A}},\alpha'_{\text{A}},\beta'_{\text{A}}}
b_{\alpha_{\text{A}},\beta_{\text{A}},\alpha'_{\text{A}},\beta'_{\text{A}}}
\left(|\alpha_{\text{A}}\rangle\langle\alpha'_{\text{A}}|\right)\left(|\tilde{\beta}_{\text{A}}\rangle\langle\tilde{\beta}'_{\text{A}}|\right), \label{eq14}
\end{align}
where
\begin{equation}
b_{\alpha_{\text{A}},\beta_{\text{A}},\alpha'_{\text{A}},\beta'_{\text{A}}}=\sum_{\gamma_{\text{B}},\gamma'_{\text{B}}} a_{\alpha_{\text{A}},\gamma_{\text{B}},\beta_{\text{A}},\gamma'_{\text{B}}} a^{*}_{\alpha'_{\text{A}},\gamma_{\text{B}},\beta'_{\text{A}},\gamma'_{\text{B}}}.\label{eq15}
\end{equation}
This yields a general formulation of the entanglement states.
The detailed derivation of Eq.(\ref{eq15}) is shown in the Appendix.
Equations (14) and (15) imply that the state in the tilde space $|\tilde{\beta}_{\text{A}}\rangle\langle\tilde{\beta}'_{\text{A}}|$ change into the state $|\alpha_{\text{A}}\rangle\langle\alpha'_{\text{A}}|$ in the original space through the fluctuation of the state $|\gamma_{\text{B}}\rangle$.
Then the parameter $b_{\alpha_{\text{A}},\beta_{\text{A}},\alpha'_{\text{A}},\beta'_{\text{A}}}$ expresses the contribution of the fluctuation of system B to system A.
This is nothing but the quantum entanglement.

In the following section, we examine some typical cases of entanglement states using the extended density matrix $\hat{\rho}_{\text{A}}$.

\section{Simple examples of two spin systems}
In this section, we apply the general formulation (\ref{eq14}) of $\hat{\rho}_{\text{A}}$ to some simple cases of two spin systems, whose Hamiltonian $\mathcal{H}$ is expressed as
\begin{equation}
\mathcal{H}=-J\Vec{S}_{\text{A}}\cdot\Vec{S}_{\text{B}}-\mu_{\text{B}}(H_{\text{A}} S_{\text{A}}^{z}+H_{\text{B}} S_{\text{B}}^{z})\label{eq16}
\end{equation}
using the spin operators $\Vec{S}_{\text{A}}=(S_{\text{A}}^x,S_{\text{A}}^y,S_{\text{A}}^z)$ and $\Vec{S}_{\text{B}}=(S_{\text{B}}^x,S_{\text{B}}^y,S_{\text{B}}^z)$.
Here, $H_{\text{A}}$ and $H_{\text{B}}$ denote the external field conjugate to $\Vec{S}_{\text{A}}$ and $\Vec{S}_{\text{B}}$, respectively.
A matrix form of the Hamiltonian $\mathcal{H}$ is obtained as
\begin{equation}
\mathcal{H}=
\begin{pmatrix}
-\frac{J}{4}-\mu_{\text{B}}\frac{H_{\text{A}}+H_{\text{B}}}{2}&0&0&0
\\
0&\frac{J}{4}-\mu_{\text{B}}\frac{H_{\text{A}}-H_{\text{B}}}{2}&-\frac{J}{2}&0
\\
0&-\frac{J}{2}&\frac{J}{4}+\mu_{\text{B}}\frac{H_{\text{A}}-H_{\text{B}}}{2}&0
\\
0&0&0&-\frac{J}{4}+\mu_{\text{B}}\frac{H_{\text{A}}+H_{\text{B}}}{2}
\end{pmatrix}\label{eq17}
\end{equation} 
using the bases $\{|++\rangle,|+-\rangle,|-+\rangle,|--\rangle\}$.
In the following subsections, we consider some typical cases of the Hamiltonian (\ref{eq17}).

\subsection{Equilibrium systems without external fields $(H_{\text{A}}=H_{\text{B}}=0)$}
In this subsection, we consider the equilibrium states with the Hamiltonian
\begin{equation}
\mathcal{H}=-J\Vec{S}_{\text{A}}\cdot\Vec{S}_{\text{B}}=-J(S_{\text{A}}^xS_{\text{B}}^x+S_{\text{A}}^yS_{\text{B}}^y+S_{\text{A}}^zS_{\text{B}}^z).\label{eq18}
\end{equation}
In this case, the ordinary density matrix $\rho_{\text{eq}}$ is obtained as follows;
\begin{align}
\rho_{\text{eq}}&=e^{-\beta \mathcal{H}}/Z(\beta)\notag
\\
&=\frac{1}{Z(\beta)}e^{-K/4}
\begin{pmatrix}
e^{K/2}&0&0&0
\\
0&\cosh K/2&\sinh K/2&0
\\
0&\sinh K/2&\cosh K/2&0
\\
0&0&0&e^{K/2}
\end{pmatrix},\label{eq19}
\end{align}
where the partition function is denoted by
\begin{equation}
Z(\beta)={\text{Tr}}e^{-\beta\mathcal{H}}=2e^{-K/4}\left( e^{K/2}+\cosh K/2 \right).\label{eq20}
\end{equation}
Here, $\beta$ and $K$ denote the inverse temperature $1/k_{\text{B}}T$ and the scaled interaction $\beta J$, respectively.
For the density matrix $\rho_{\text{eq}}$, we obtain $\rho_{\text{eq}}^{1/2}$ as
\begin{equation}
\rho_{\text{eq}}^{1/2}=
\begin{pmatrix}
\frac{1}{\sqrt{3+e^{-K}}}&0&0&0
\\
0&\frac{1}{2}\left( \frac{1}{\sqrt{3+e^{-K}}} + \frac{1}{\sqrt{1+3e^{K}}}\right) &\frac{1}{2}\left( \frac{1}{\sqrt{3+e^{-K}}} - \frac{1}{\sqrt{1+3e^{K}}} \right)&0
\\
0&\frac{1}{2}\left( \frac{1}{\sqrt{3+e^{-K}}} - \frac{1}{\sqrt{1+3e^{K}}}\right)&\frac{1}{2}\left( \frac{1}{\sqrt{3+e^{-K}}} + \frac{1}{\sqrt{1+3e^{K}}}\right)&0
\\
0&0&0&\frac{1}{\sqrt{3+e^{-K}}}
\end{pmatrix},\label{eq21}
\end{equation}
from the relation
\begin{equation}
(\rho_{\text{eq}}^{1/2})^2=\rho_{\text{eq}}.\label{eq22}
\end{equation}
Now, the variables of the spin B, namely $\Vec{S}_{\text{B}}$, are assumed as hidden variables.
Then the formulations (\ref{eq14}) and (\ref{eq15}) give the extended density matrix $\hat{\rho}_{\text{A}}$ of the spin A as
\begin{align}
\hat{\rho}_{\text{A}}=&b_{\text{d1}} |\+ \tilde{\+} \rangle \langle \+ \tilde{\+}| \quad + \quad b_{\text{d2}} |\- \tilde{\-} \rangle \langle \- \tilde{\-}|\notag
\\
&+b_{\text{cf}}\left( |\+ \tilde{\+} \rangle \langle \- \tilde{\-}| \quad + \quad |\- \tilde{\-} \rangle \langle \+ \tilde{\+}| \right)\notag
\\
&+b_{\text{qe}}\left( |\+ \tilde{\-} \rangle \langle \+ \tilde{\-}| \quad + \quad |\- \tilde{\+} \rangle \langle \- \tilde{\+}| \right)\notag
\\
=&b_{\text{d}}\left( |\+\rangle\langle \+||\tilde{\+}\rangle\langle \tilde{\+}| \quad + \quad |\-\rangle\langle \-||\tilde{\-}\rangle\langle \tilde{\-}| \right)\notag
\\
&+b_{\text{cf}}\left( |\+\rangle\langle \-||\tilde{\+}\rangle\langle \tilde{\-}| \quad + \quad |\-\rangle\langle \+||\tilde{\-}\rangle\langle \tilde{\+}| \right)\notag
\\
&+b_{\text{qe}}\left( |\+\rangle\langle \+||\tilde{\-}\rangle\langle \tilde{\-}| \quad + \quad |\-\rangle\langle \-||\tilde{\+}\rangle\langle \tilde{\+}| \right).\label{eq23}
\end{align}
Here the coefficients $b_{\text{d}}(=b_{\text{d1}}=b_{\text{d2}}),b_{\text{cf}}$ and $b_{\text{qe}}$ are obtained by Eq.(\ref{eq15}) as follows:
\begin{align}
b_{\text{d}}&\equiv b_{++++}(=b_{----})=\sum_{\gamma_{\text{B}},\gamma'_{\text{B}}}a_{+\gamma_{\text{B}}+\gamma'_{\text{B}}}a^{*}_{+\gamma_{\text{B}}+\gamma'_{\text{B}}}\notag
\\
&=a_{++++}a^{*}_{++++}+a_{+++-}a^{*}_{+++-}+a_{+-++}a^{*}_{+-++}+a_{+-+-}a^{*}_{+-+-}\notag
\\
&=\left(\frac{1}{\sqrt{3+e^{-K}}}\right)^2+0+0+\frac{1}{4}\left( \frac{1}{\sqrt{3+e^{-K}}} + \frac{1}{\sqrt{1+3e^{K}}}\right)^2\notag
\\
&=\frac{1}{3+e^{-K}}+\frac{1}{4}\left( \frac{1}{\sqrt{3+e^{-K}}} + \frac{1}{\sqrt{1+3e^{K}}}\right)^2,
\label{eq24}
\end{align}
\begin{align}
b_{\text{cf}}&\equiv b_{++--}(=b_{--++})=\sum_{\gamma_{\text{B}},\gamma'_{\text{B}}}a_{+\gamma_{\text{B}}+\gamma'_{\text{B}}}a^{*}_{-\gamma_{\text{B}}-\gamma'_{\text{B}}}\notag
\\
&=a_{++++}a^{*}_{-+-+}+a_{+++-}a^{*}_{-+--}+a_{+-++}a^{*}_{---+}+a_{+-+-}a^{*}_{----}\notag
\\
&=\frac{1}{3+e^{-K}}+\frac{1}{\sqrt{(3+e^{-K})(1+3e^{K})}},\label{eq25}
\end{align}
and
\begin{align}
b_{\text{qe}}&\equiv b_{+-+-}(=b_{-+-+})=\sum_{\gamma_{\text{B}},\gamma'_{\text{B}}}a_{+\gamma_{\text{B}}-\gamma'_{\text{B}}}a^{*}_{+\gamma_{\text{B}}-\gamma'_{\text{B}}}\notag
\\
&=a_{++-+}a^{*}_{++-+}+a_{++--}a^{*}_{++--}+a_{+--+}a^{*}_{+--+}+a_{+---}a^{*}_{+---}\notag
\\
&=\frac{1}{4}\left( \frac{1}{\sqrt{3+e^{-K}}} - \frac{1}{\sqrt{1+3e^{K}}} \right)^2,\label{eq26}
\end{align}
where the elements $a_{++++},a_{+-+-},\cdots$ denote the matrix elements of the density matrix $\rho^{1/2}_{\text{eq}}$ in Eq.(\ref{eq21}).
The above parameters $b_{\text{qe}},b_{\text{cf}}$ and $b_{\text{d}}$ correspond to the quantum entanglements (qe), the classical fluctuations (cf) and the diagonal components (d) of the extended density matrix $\hat{\rho}_{\text{A}}$, respectively.
The reason for such naming will be explained below.

\begin{figure}[hbpt]
\begin{center}
\includegraphics[width=7cm]{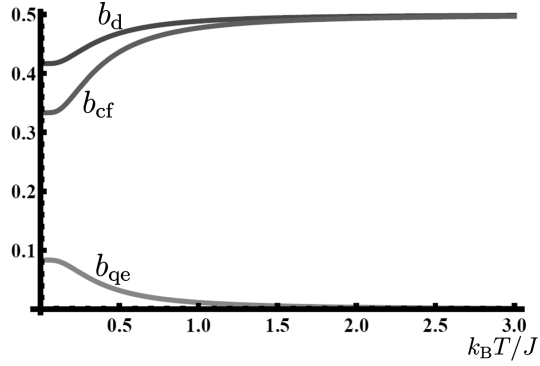}
\end{center}
\caption{Temperature dependence of the parameters $b_{\text{d1}}, b_{\text{cf}}, b_{\text{qe}}$ and $b_{\text{d2}}$, which are given analytically by Eqs.(\ref{eq24}), (\ref{eq25}) and (\ref{eq26}), respectively. The horizontal axis expresses the scaled temperature $k_{\text{B}}T/J$. Here the equality $b_{\text{d1}}=b_{\text{d2}}$ holds, because the present system expressed by Hamiltonian (\ref{eq18}) has the symmetry for the spin inversion. The quantum entanglement described by $b_{\text{qe}}$ vanishes at high temperatures.}
\label{fig1}
\end{figure}

As the variables of the spin B are traced out, the states in Eq.(\ref{eq23}) express the variables of the spin A.
For example, the first state symbol $|\+\rangle\langle \+||\tilde{\+}\rangle\langle \tilde{\+}| $ (and $|\-\rangle\langle \-||\tilde{\-}\rangle\langle \tilde{\-}| $) in Eq.(\ref{eq23}) means that the spin A takes the up state (and down state) both in the original space and in the tilde space.
The second state symbols $|\+\rangle\langle \-||\tilde{\+}\rangle\langle \tilde{\-}| $ and $|\-\rangle\langle \+||\tilde{\-}\rangle\langle \tilde{\+}| $ mean the classical fluctuations.
That is, the original state $|\+\rangle\langle \-| $ is combined with the same type of tilde state $|\tilde{\+}\rangle\langle \tilde{\-}| $.
(Similarly, we have the combined state $|\-\rangle\langle \+||\tilde{\-}\rangle\langle \tilde{\+}| $.)
As shown in Fig.\ref{fig1}, the parameter $b_{\text{cf}}$ (corresponding to the ``classical fluctuation") monotonically increases as the temperature increases. 
This classical fluctuation is caused by the thermal fluctuation.
It appears even in such classical systems as the Ising model, the classical ideal gas, and the Debye model. 
On the other hand, the symbol $|\+\rangle\langle \+||\tilde{\-}\rangle\langle \tilde{\-}| $ means that the spin A takes the state $|\+\rangle\langle \+|$ in the original space different from the state $|\tilde{\-}\rangle\langle \tilde{\-}| $ in the tilde space.
Then the parameter $b_{\text{qe}}$ expresses the effect of ``quantum entanglement".
This quantum fluctuation appears only in quantum systems, and it is used as an order parameter of quantum systems at zero temperature in many cases [17-22].
The temperature dependences of $b_{\text{d1}},b_{\text{cf}},b_{\text{qe}}$ and $b_{\text{d2}}$ are shown in Fig.\ref{fig1}.
It is easily seen that the quantum entanglement denoted by $b_{\text{qe}}$ vanishes for high temperatures.
From the above discussion, only the intrinsic quantum entanglement is extracted clearly in this formulation based on the TFD.
In particular, we can understand the entangled state of the system through such a ``single product" as $|\+\rangle\langle \+||\tilde{\-}\rangle\langle \tilde{\-}|$ through the extended density matrix $\hat{\rho}_{\text{A}}$.

\subsection{Non-equilibrium systems on the ground states $(H_{\text{A}}=H_{\text{B}}=0$ and $T=0)$}
\begin{figure}[hbpt]
\begin{center}
\includegraphics[width=7cm]{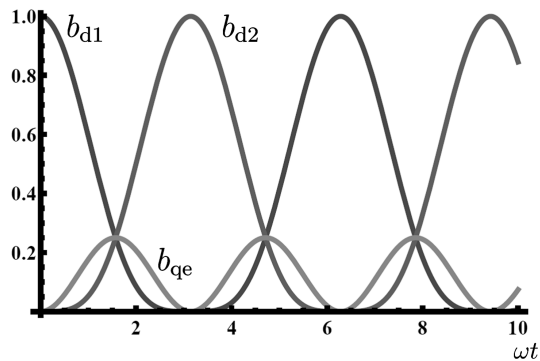}
\end{center}
\caption{Time dependence of the parameters $b_{\text{d1}}, b_{\text{qe}}$ and $b_{\text{d2}}$ on the non-dissipative system, (which are defined in Eq.(\ref{eq40})). The horizontal axis expresses the scaled time $\omega t$. This finite system shows an oscillatory behavior. When the parameter $b_{\text{d1}}$ or $b_{\text{cf}}$ takes larger values, the parameter $b_{\text{qe}}$ takes a smaller value. This shows a kind of classical-quantum crossover-oscillations.}
\label{fig2} 
\end{figure}

Now we consider the Hamiltonian (\ref{eq18}) as a non-equilibrium system in the ground states.
The time dependence of the ordinary density matrix $\rho(t)$ is given by the von Neumann equation;
\begin{equation}
i\hbar\frac{\partial}{\partial t}\rho(t)=[\mathcal{H},\rho(t)].\label{eq27}
\end{equation}
The solution of Eq.(\ref{eq27}) is given in the form
\begin{equation}
\rho(t)=U^{\dagger}(t)\rho_0 U(t)\label{eq28}
\end{equation}
for the initial condition $\rho_0$, where the unitary operator $U(t)$ denotes
\begin{equation}
U(t)=e^{i\mathcal{H}t/\hbar}.\label{eq29}
\end{equation}
Then the density matrix $\rho^{1/2}(t)$ is given in the form
\begin{equation}
\rho^{1/2}(t)=U^{\dagger}(t)\rho_0^{1/2} U(t).\label{eq30}
\end{equation} 
From Eq.(\ref{eq29}), the unitary operator $U(t)$ is expressed in the matrix form as
\begin{equation}
U(t)=e^{i\mathcal{H}t/\hbar}=e^{i\omega t/4}
\begin{pmatrix}
e^{-i\omega t/2}&0&0&0
\\
0&\cos (\omega t/2)&-i\sin (\omega t/2)&0
\\
0&-i\sin (\omega t/2)&\cos (\omega t/2)&0
\\
0&0&0&e^{-i\omega t/2}
\end{pmatrix}\label{eq31}
\end{equation}
using the bases $\{|++\rangle,|+-\rangle,|-+\rangle,|--\rangle\}$, where the parameter $\omega$ is defined as $\omega\equiv J/\hbar$.
From Eqs.(\ref{eq30}) and (\ref{eq31}), we obtain the time-dependent density matrix $\rho^{1/2}(t)$ as follows:
\begin{align}
\rho^{1/2}(t)&=U^{\dagger}(t)\rho_0^{1/2} U(t)=U^{\dagger}(t)
\begin{pmatrix}
P_{++}^{1/2}&0&0&0
\\
0&P_{+-}^{1/2}&0&0
\\
0&0&P_{-+}^{1/2}&0
\\
0&0&0&P_{--}^{1/2}
\end{pmatrix}
U(t)\notag
\\
&=
\begin{pmatrix}
P_{++}^{1/2}&0&0&0
\\
0&\frac{1}{2}\left(P_{+}^{1/2}+P_{-}^{1/2}\cos \omega t\right) &-\frac{i}{2}P_{-}^{1/2}\sin \omega t &0
\\
0&\frac{i}{2}P_{-}^{1/2}\sin \omega t &\frac{1}{2}\left(P_{+}^{1/2}-P_{-}^{1/2}\cos \omega t\right) &0
\\
0&0&0&P_{--}^{1/2} 
\end{pmatrix},\label{eq32}
\end{align}
for the initial condition
\begin{equation}
\rho_0=
\begin{pmatrix}
P_{++}&0&0&0
\\
0&P_{+-}&0&0
\\
0&0&P_{-+}&0
\\
0&0&0&P_{--}
\end{pmatrix}.\label{eq33}
\end{equation}
Here the parameters $P_{+}^{1/2}$ and $P_{-}^{1/2}$ are defined by
\begin{equation}
P_{+}^{1/2}\equiv P_{+-}^{1/2}+P_{-+}^{1/2}{\text{ and }}P_{-}^{1/2}\equiv P_{+-}^{1/2}-P_{-+}^{1/2}.\label{eq34}
\end{equation}
Then, using the formulation (\ref{eq14}) and such non-zero elements of $\rho^{1/2}(t)$ as shown in Eq.(\ref{eq32}), we obtain the extended density matrix $\hat{\rho}_A$ of the spin A in the form
\begin{align}
\hat{\rho}_{\text{A}}(t)=&b_{++++}|\+\rangle\langle \+||\tilde{\+}\rangle\langle \tilde{\+}| \quad + \quad b_{+-+-}|\+\rangle\langle \+||\tilde{\-}\rangle\langle \tilde{\-}|\notag
\\
&+b_{++--}|\+\rangle\langle \-||\tilde{\+}\rangle\langle \tilde{\-}| \quad + \quad b_{--++}|\-\rangle\langle \+||\tilde{\-}\rangle\langle \tilde{\+}|\notag
\\
&+b_{-+-+}|\-\rangle\langle \-||\tilde{\+}\rangle\langle \tilde{\+}| \quad + \quad b_{----}|\-\rangle\langle \-||\tilde{\-}\rangle\langle \tilde{\-}|.\label{eq35}  
\end{align}
Here the coefficients $b_{++++},b_{++--},b_{+-+-},b_{-+-+},b_{--++}$ and $b_{----}$ are obtained by Eq.(\ref{eq15}) as follows:
\begin{align}
b_{++++}=&P_{++}+\frac{1}{4}\left( P_{+}^{1/2}+P_{-}^{1/2}\cos \omega t\right)^2,\label{eq36}
\\
b_{++--}=b_{--++}=&\frac{1}{2}P_{++}^{1/2}\left( P_{+}^{1/2}-P_{-}^{1/2}\cos \omega t\right)\notag
\\
&+\frac{1}{2}P_{--}^{1/2}\left( P_{+}^{1/2}+P_{-}^{1/2}\cos \omega t\right),\label{eq37}
\\
b_{+-+-}=b_{-+-+}=&\frac{1}{4}P_{-}\sin^2 \omega t,\label{eq38}
\\
b_{----}=&P_{--}+\frac{1}{4}\left( P_{+}^{1/2}-P_{-}^{1/2}\cos \omega t\right)^2.\label{eq39}
\end{align}
These are still complicated and it is difficult to understand the physical meanings.
Thus we assume the initial condition that $P_{++}=P_{--}=P_{-+}=0$ and $P_{+-}=1$.
This initial condition means the classical condition, namely spin A takes up and spin B takes down.
Then the extended density matrix of the spin A (as shown in Eq.(\ref{eq35})) is simply expressed as
\begin{align}
\hat{\rho}_{\text{A}}(t)=&b_{\text{d1}}|\+\rangle\langle \+||\tilde{\+}\rangle\langle \tilde{\+}| \quad + \quad b_{\text{d2}}|\-\rangle\langle \-||\tilde{\-}\rangle\langle \tilde{\-}|\notag
\\
&+b_{\text{qe}}\left( |\+\rangle\langle \+||\tilde{\-}\rangle\langle \tilde{\-}| \quad + \quad |\-\rangle\langle \-||\tilde{\+}\rangle\langle \tilde{\+}|  \right)\notag
\\
=& \cos^4\frac{\omega t}{2}|\+\rangle\langle \+||\tilde{\+}\rangle\langle \tilde{\+}| \quad + \quad \sin^4 \frac{\omega t}{2}|\-\rangle\langle \-||\tilde{\-}\rangle\langle \tilde{\-}|\notag
\\
&+\frac{1}{4}\sin^2 \omega t\left( |\+\rangle\langle \+||\tilde{\-}\rangle\langle \tilde{\-}| \quad + \quad |\-\rangle\langle \-||\tilde{\+}\rangle\langle \tilde{\+}|  \right).\label{eq40}
\end{align}
Here, the parameters $b_{\text{d1}},b_{\text{qe}}$ and $b_{\text{d2}}$ correspond to those in Eq.(\ref{eq23}).
The extended density matrix $\hat{\rho}_{\text{A}}(t)$ in Eq.(\ref{eq40}) contains physical information on quantum entanglement.
The state symbol $|\+\rangle\langle \+||\tilde{\+}\rangle\langle \tilde{\+}|$ means the up states of spin A both in the original and tilde spaces, while the state symbol $|\-\rangle\langle \-||\tilde{\-}\rangle\langle \tilde{\-}|$ means the down state, as shown in the previous subsection.
When either of them takes a dominant value, namely $t=n\pi/\omega$ for $n=0,1,2,\cdots$, the spin A takes a classical state, up or down.
In contrast, the state symbols $|\+\rangle\langle \+||\tilde{\-}\rangle\langle \tilde{\-}|$ and $|-\rangle\langle -||\tilde{+}\rangle\langle \tilde{+}|$ mean the entangled states.
If it takes a dominant value, namely $t=(2n+1)\pi/(2\omega)$, the spin A takes a quantum entangled state.
This crossover-oscillation between classical and quantum states is shown in Fig.{\ref{fig2}}.
Of course, in this case, the behavior of the spin A shows the oscillatory time-dependence because we have considered the finite system of two spins.
When the parameter $b_{\text{d1}}$ or $b_{\text{cf}}$ takes larger values, the parameter $b_{\text{qe}}$ takes smaller values.
This shows a kind of classical-quantum crossover-oscillations as shown in Fig.\ref{fig2}.

\subsection{Non-equilibrium systems with heat bath $(H_{\text{A}}=H_{\text{B}}=0)$}
\begin{figure}[hbpt]
\begin{center}
\includegraphics[width=7cm]{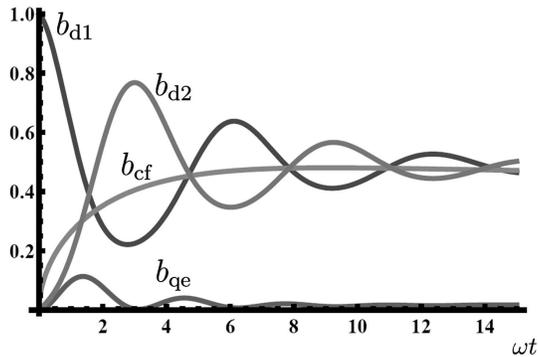}
\end{center}
\caption{Time dependence of the parameters $b_{\text{d1}},b_{\text{cf}},b_{\text{qe}}$ and $b_{\text{d2}}$ in the dissipative system, (which are defined by Eq.(\ref{eq43})).  The horizontal axis expresses the scaled time $\omega t$. We have used the scaled dissipation rate $\epsilon /\omega =0.2$ and the scaled temperature $k_{\text{B}}T/J=0.7$.}
\label{fig3}
\end{figure}

In the previous section, we have discussed the system of a finite size and consequently without dissipative mechanisms.
To include dissipative mechanism, we consider the dissipative von Neumann equation \cite{26,27}
\begin{equation}
i\hbar\frac{\partial}{\partial t}\rho(t)=[\mathcal{H},\rho(t)]-\epsilon \left( \rho(t)-\rho_{\text{eq}} \right),\label{eq41}
\end{equation}
in the present subsection.
The solution of Eq.(\ref{eq41}) is given by
\begin{equation}
\rho(t)=e^{-\epsilon t}U^{\dagger}(t)\rho_0 U(t)+(1-e^{-\epsilon t})\rho_{\text{eq}}.\label{eq42}
\end{equation}
Similarly to the previous manipulation, the extended density matrix $\hat{\rho}_{\text{A}}(t)$ is obtained as
\begin{align}
\hat{\rho}_A(t)=&b_{\text{d1}}|\+\rangle\langle \+||\tilde{\+}\rangle\langle \tilde{\+}| \quad + \quad b_{\text{d2}}|\-\rangle\langle \-||\tilde{\-}\rangle\langle \tilde{\-}|\notag
\\
&+b_{\text{cf}}\left( |\+\rangle\langle \-||\tilde{\+}\rangle\langle \tilde{\-}| \quad + \quad |\-\rangle\langle \+||\tilde{\-}\rangle\langle \tilde{\+}| \right)\notag
\\
&+b_{\text{qe}}\left( |\+\rangle\langle \+||\tilde{\-}\rangle\langle \tilde{\-}| \quad + \quad |\-\rangle\langle \-||\tilde{\+}\rangle\langle \tilde{\+}|  \right)\label{eq43}
\end{align}
for the initial condition $P_{++}=P_{--}=P_{-+}=0$ and $P_{+-}=1$.
To derive Eq.(\ref{eq43}), we have used Eqs.(\ref{eq19}) and (\ref{eq32}) for $\rho_{\text{eq}}$ and $U^{\dagger}(t)\rho_0 U(t)$, respectively.
The parameters $b_{\text{d1}},b_{\text{cf}},b_{\text{qe}}$ and $b_{\text{d2}}$ are obtained as functions of $t,\epsilon,\omega,T$ and $J$ analytically.
However, they are too complicated and it is difficult to understand the physical meaning.
Thus, we try to show their numerical behaviors in Fig.\ref{fig3}.
As shown in Fig.\ref{fig3}, the parameters $b_{\text{d1}}$ and $b_{\text{d2}}$ describing the classical oscillation approach the same value $b_{\text{d1}}=b_{\text{d2}}$ for $t\to\infty$.
Consequently, the parameter $b_{\text{qe}}$ describing the quantum oscillation approaches the equilibrium value $b_{\text{qe}}$ given by Eq.(\ref{eq26}).
As discussed in subsection 3.1, the parameter $b_{\text{cf}}$ expresses the thermal but classical fluctuations. 

The classically fluctuating states $|\+\rangle\langle \-||\tilde{\+}\rangle\langle \tilde{\-}|$ and $|\-\rangle\langle \+||\tilde{\-}\rangle\langle \tilde{\+}|$, namely the parameter $b_{\text{cf}}$, appear both in the thermal equilibrium system (as shown in Fig.\ref{fig1}) and in the dissipative system (as shown in Fig.\ref{fig3}), but they do not appear in the non-dissipative system as shown in Fig.\ref{fig2}.
The above three examples show that our method using the extended density matrix is useful in distinguishing the fluctuations based on the quantum entanglement from the thermal but classical fluctuations. 

\subsection{Frustration effect with competitive external fields $(H_{\text{A}}=H_{\text{B}}=H$ and $H_{\text{A}}=-H_{\text{B}}=H)$}
\begin{figure}[hbpt]
\begin{center}
\includegraphics[width=12cm]{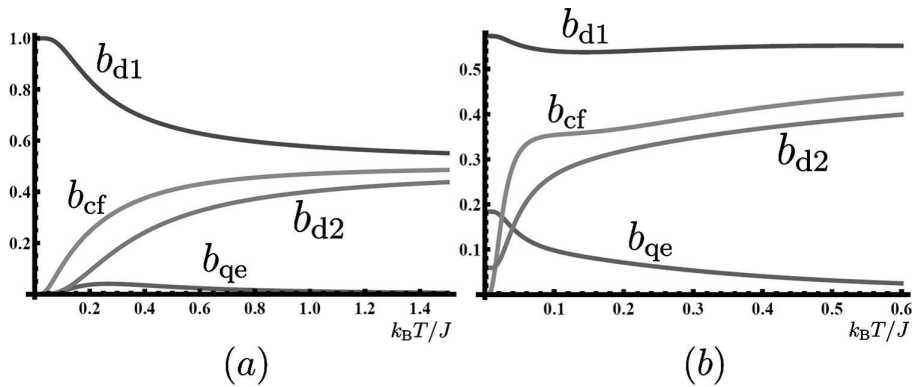}
\end{center}
\caption{Temperature dependence of the parameters $b_{\text{d1}},b_{\text{cf}},b_{\text{qe}}$ and $b_{\text{d2}}$ given by Eq.(\ref{eq50}) in the presence of an external field ($\mu_{\text{B}}H/J=0.3$). Figure (a) shows their numerical results in the non-competitive model (\ref{eq46}), while figure (b) shows their numerical results in the competitive model (\ref{eq47}). At the ground state ($T=0$), the parameter $b_{\text{qe}}$ becomes zero in the non-competitive model, but it is non-vanishing in the competitive model.}
\label{fig4}
\end{figure}

We study here the frustration effect for the entanglement.
For this purpose, we compare the following two systems
\begin{equation}
\mathcal{H}^{\text{noncomp.}}=-J\Vec{S}_{\text{A}}\cdot\Vec{S}_{\text{B}}-\mu_{\text{B}}H(S_{\text{A}}^z+S_{\text{B}}^z)\label{eq44}
\end{equation}
and
\begin{equation}
\mathcal{H}^{\text{comp.}}=-J\Vec{S}_{\text{A}}\cdot\Vec{S}_{\text{B}}-\mu_{\text{B}}H(S_{\text{A}}^z-S_{\text{B}}^z).\label{eq45}
\end{equation}
The Hamiltonian $\mathcal{H}^{\text{noncomp.}}$ does not contain a competition effect among the interaction $J\Vec{S}_{\text{A}}\cdot\Vec{S}_{\text{B}}$ and the external fields $\mu_{\text{B}}H(S_{\text{A}}^z+S_{\text{B}}^z)$, while the Hamiltonian $\mathcal{H}^{\text{comp.}}$ contains the competition effect.
In the present study, we describe this competition as a kind of frustration.
For these assumptions, $\mathcal{H}^{\text{noncomp.}}$ and $\mathcal{H}^{\text{comp.}}$ are expressed in the matrix forms as follows:
\begin{equation}
\mathcal{H}^{\text{noncomp.}}=
\begin{pmatrix}
-\frac{J}{4}-\mu_{\text{B}}H&0&0&0
\\
0&\frac{J}{4}&-\frac{J}{2}&0
\\
0&-\frac{J}{2}&\frac{J}{4}&0
\\
0&0&0&-\frac{J}{4}+\mu_{\text{B}}H
\end{pmatrix},\label{eq46}
\end{equation}
and
\begin{equation}
\mathcal{H}^{\text{comp.}}=
\begin{pmatrix}
-\frac{J}{4}&0&0&0
\\
0&\frac{J}{4}-\mu_{\text{B}}H&-\frac{J}{2}&0
\\
0&-\frac{J}{2}&\frac{J}{4}+\mu_{\text{B}}H&0
\\
0&0&0&-\frac{J}{4}
\end{pmatrix}.\label{eq47}
\end{equation}
Then the density matrices $\rho_{\text{eq}}^{\text{noncomp.}}$ and $\rho_{\text{eq}}^{\text{comp.}}$ are obtained as
\begin{equation}
\rho_{\text{eq}}^{\text{noncomp.}}=\frac{1}{Z^{\text{noncomp.}}(\beta)}
\begin{pmatrix}
e^{K/4+h}&0&0&0
\\
0&e^{-K/4}\cosh\frac{K}{2}&e^{-K/4}\sinh\frac{K}{2}&0
\\
0&e^{-K/4}\sinh\frac{K}{2}&e^{-K/4}\cosh\frac{K}{2}&0
\\
0&0&0&e^{K/4-h}
\end{pmatrix},\label{eq48}
\end{equation}
and
\begin{equation}
\rho_{\text{eq}}^{\text{comp.}}=\frac{1}{Z^{\text{comp.}}(\beta)}
\begin{pmatrix}
e^{K/4}&0&0&0
\\
0&e^{-K/4}\left(\cosh{L} +\frac{h}{{L}}\sinh{L} \right) & \frac{K}{2{L}}e^{-K/4}\sinh{L} &0
\\
0&\frac{K}{2{L}}e^{-K/4}\sinh{L} &e^{-K/4}\left(\cosh{L} -\frac{h}{{L}}\sinh{L} \right) &0
\\
0&0&0&e^{K/4}
\end{pmatrix},\label{eq49}
\end{equation}
respectively.
Here the parameters $h$ and $L$ denote $h=\beta \mu_{\text{B}}H$ and $L=\sqrt{h^2+K^2/4}$, respectively.
Then we can obtain the extended density matrix $\hat{\rho}_{\text{A}}$ as
\begin{align}
\hat{\rho}_\text{A}^{\alpha}(t)=&b_{\text{d1}}^{\alpha}|\+\rangle\langle \+||\tilde{\+}\rangle\langle \tilde{\+}| \quad + \quad b_{\text{d2}}^{\alpha}|\-\rangle\langle \-||\tilde{\-}\rangle\langle \tilde{\-}|\notag
\\
&+b_{\text{cf}}^{\alpha}\left( |\+\rangle\langle \-||\tilde{\+}\rangle\langle \tilde{\-}| \quad + \quad |\-\rangle\langle \+||\tilde{\-}\rangle\langle \tilde{\+}| \right)\notag
\\
&+b_{\text{qe}}^{\alpha}\left( |\+\rangle\langle \+||\tilde{\-}\rangle\langle \tilde{\-}| \quad + \quad |\-\rangle\langle \-||\tilde{\+}\rangle\langle \tilde{\+}|  \right),\label{eq50}
\end{align}
where $\alpha$ denotes ``noncomp." or ``comp."
The behaviors of parameters $\{b_i^{\alpha}\}$ $(i=1,\cdots,4)$ are shown numerically in Fig.\ref{fig4}.
The figure \ref{fig4}-(a) shows the parameters $\{b_i^{\text{noncomp.}}\}$ while the figure \ref{fig4}-(b) shows $\{b_i^{\text{comp.}}\}$.
In the non-competitive system, the external field $H$ breaks the symmetry of the spin inversion and the parameter $b_{\text{d1}}^{\text{noncomp.}}=1$ for $T=0$, because the entanglement parameter $b_{\text{qe}}^{\text{noncomp.}}=0$ for $T=0$.
On the other hand, in the competitive system, the frustration makes a finite entanglement ($b_{\text{qe}}^{\text{comp.}}\not=0$) for $T=0$, even under the finite external field $H$.
Thus the parameter $b_{\text{d1}}^{\text{comp.}}$ which expresses the probability weight of the up state is smaller than the maximum $1.0$.
This is a typical example of the entanglement caused by the frustration.

\section{Correspondence of the entanglement entropy and the parameter $b_{\text{qe}}$} 
From the above examples, it may be clarified that the transition term between the original space and the tilde space (i.e. $ |\+\rangle\langle \+||\tilde{\-}\rangle\langle \tilde{\-}| $ or $|\-\rangle\langle \-||\tilde{\+}\rangle\langle \tilde{\+}|$) expresses the strength of the entanglement.
In this section, we verify the correspondence of the parameter $b_{\text{qe}}$ and the entanglement entropy $S$ for the simple case discussed in Section 3.2.
The entanglement entropy in our case is defined by Eq.(\ref{eq6}).
The ordinary density matrix $\rho(t)$ is obtained from Eq.(\ref{eq32}) for the Hamiltonian (\ref{eq18}) in this non-dissipative system.
Here we assume that the initial condition $P_{++}=P_{--}=P_{-+}=0$ and $P_{+-}=1$.
Then the density matrix $\rho(t)$ is obtained as
\begin{align}
\rho(t)=&\cos^2\left(\frac{\omega t}{2}\right)|\+ \- \rangle\langle \+ \-| \quad - \quad \frac{i}{2}\sin(\omega t)|\+ \-\rangle\langle \- \+|\notag
\\
&+\frac{i}{2}\sin(\omega t)|\- \+ \rangle\langle \+ \-| \quad + \quad \sin^2\left(\frac{\omega t}{2}\right)|\- \+\rangle\langle \- \+|.\label{eq51}
\end{align}
Then $\rho_{\text{A}}$ in Eq.(\ref{eq6}) is derived as
\begin{align}
\rho_{\text{A}} &={\text{Tr}}_{\text{B}}\rho(t)=_{\text{B}}\!\!\langle \+|\rho(t)|\+ \rangle_{\text{B}} \quad + \quad _{\text{B}}\langle \-|\rho(t)|\- \rangle_{\text{B}}\notag
\\
&=\cos^2\left(\frac{\omega t}{2}\right)|\+\rangle\langle \+| \quad + \quad \sin^2\left(\frac{\omega t}{2}\right)|\-\rangle\langle \-|,\label{eq52}
\end{align}
where $|\+\rangle_{\text{B}}$ and $|\-\rangle_{\text{B}}$ correspond to the spin states of the spin B, and $|\+\rangle$ and $|\-\rangle$ correspond to the spin states of the spin A.
Thus the entanglement entropy $S$ is obtained as
\begin{equation}
S=-k_{\text{B}}\left[ \cos^2\left(\frac{\omega t}{2}\right)\log \left( \cos^2\left(\frac{\omega t}{2}\right) \right) +\sin^2\left(\frac{\omega t}{2}\right)\log \left( \sin^2\left(\frac{\omega t}{2}\right) \right) \right], \label{eq53}
\end{equation}
using Eq.(\ref{eq6}).

\begin{figure}[hbpt]
\begin{center}
\includegraphics[width=7cm]{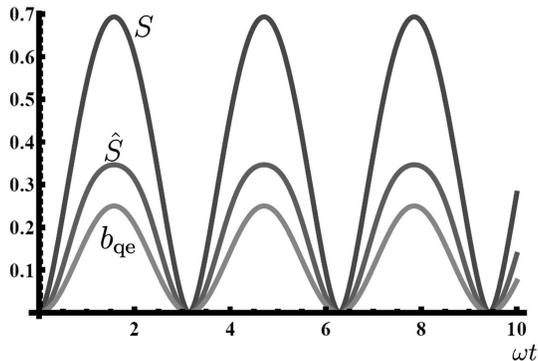}
\end{center}
\caption{Time dependence of the two kinds of entropy $S,\hat{S}$ and the parameter $b_{\text{qe}}$ in the non-dissipative system described by Eq.(\ref{eq18}). All the curves of $S,\hat{S}$ and $b_{\text{qe}}$ (showing the entanglement) have the same phase. However, their amplitudes change periodically. As the entropy $S$ (namely Eq.(\ref{eq6})) includes the classical fluctuations of the spin A, the parameter $S$ is larger than $\hat{S}$, namely $S>\hat{S}$.}
\label{fig5}
\end{figure}

Now, we try to define the ``extended" entanglement entropy using the extended density matrix $\hat{\rho}_A$ as
\begin{equation}
\hat{S}=-k_{\text{B}}{\text{Tr}}\hat{\rho}_A\log \hat{\rho}_A.\label{eq54}
\end{equation}
Here the extended density matrix $\hat{\rho}_{\text{A}}$ is given by Eq.(\ref{eq40}).
One non-zero eigenvalue of $\hat{\rho}_{\text{A}}$ is easily obtained as $(3+\cos (2\omega t))/4$.
Thus, the ``extended" entanglement entropy $\hat{S}$ yields
\begin{equation}
\hat{S}=-\frac{k_{\text{B}}}{4}(3+\cos (2\omega t))\log \left(\frac{3+\cos (2\omega t)}{4}\right).\label{eq55}
\end{equation}
The time dependences of $S, \hat{S}$ and $b_{\text{qe}}$ are shown in Fig.\ref{fig5}.
In the present finite size system, the entanglement shows the periodic oscillation.
As shown in Fig.\ref{fig5}, all the curves $S,\hat{S}$ and $b_{\text{qe}}$ showing the entanglement have the same phase.
However, their amplitudes are different from each other.
Especially, the traditional entanglement entropy $S$ is larger than the extended entanglement entropy $\hat{S}$.
This is because the definition of $S$ in Eq.(\ref{eq6}) includes not only the fluctuation caused by the entanglement but also the original (statistical) fluctuation of the spin A.
Thus, our new definition of entanglement based on the TFD is more physical.

\section{Summary and discussion}
In the present study, we have introduced a new method to study quantum entanglement using thermo field dynamics.
The extended density matrix $\hat{\rho}$ including the state in the tilde space enables us to study the entanglement intuitively, because the TFD state vector shows ``dynamical" states of quantum systems.
These ``dynamically" entangled states are presented in our formulations (\ref{eq14}) and (\ref{eq15}).
In the previous sections, we demonstrate some typical cases of entanglement states using our formulations.
Our methods can be applied to larger quantum systems, and it may clarify the mechanisms of quantum entanglement in such large quantum systems, if the ordinary density matrix $\rho$ is given.
Additionally, our formulation requires, at most, the same computational time to diagonalize the ordinary density matrix, although the extended density matrix is defined in the double Hilbert space.

Furthermore, the extended entanglement entropy $\hat{S}$ is introduced using the extended density matrix $\hat{\rho}_{\text{A}}$ and it is compared with the traditional entanglement entropy $S$ (and $b_{\text{qe}}$).
From this discussion, the condition to give the maximum entangled state can be obtained by these two entropies.
Only amplitudes are different from each other.
Thus, the parameter $b_{\text{qe}}$ may be useful to study the entanglement, because it does not require the non-linear calculations such as logarithms.

The present method enables us to distinguish clearly the various states of quantum systems. 
For example, the parameters $b_{\text{d1}}$ and $b_{\text{d2}}$ correspond to the classical state as shown in Section 3.
On the other hand, the parameters $b_{\text{cf}}$ and $b_{\text{qe}}$ correspond to the fluctuations from the thermal but classical fluctuations and to the fluctuations from the quantum entanglement, respectively.

Finally, we would like to remark that the general representation theorem \cite{4,5} makes it very convenient to study the entanglement using the TFD, because this theorem ensures the correspondence between the traditional density matrix (or entanglement entropy) and the extended density matrix (or extended entanglement entropy) as shown in Eq.(\ref{eq9}).
Then the intrinsically entangled states of quantum systems are understood through such a single product as $|\+\rangle\langle \+||\tilde{\-}\rangle\langle \tilde{\-}|$.

\section*{Acknowledgments}
One of the authors (Y.H.) would like to thank Dr. H. Matsueda and Dr. O. Araki for useful discussions.

\appendix
\section{Derivation of Eqs.(\ref{eq14}) and (\ref{eq15})}
In this appendix, we show the derivation of Eqs.(\ref{eq14}) and (\ref{eq15}).
The extended density matrix $\hat{\rho}$ is defined by $\hat{\rho}=|\Psi \rangle \langle \Psi |$ as shown in Eq.(\ref{eq7}).
Then, inserting the TFD state vector as shown in Eq.(\ref{eq11}) to the extended density matrix $\hat{\rho}$, we derive as follows;
\begin{align}
\hat{\rho}&=|\Psi\rangle\langle\Psi|\notag
\\
&=\sum_{s_{\text{A}}}\sum_{s_{\text{B}}}\sum_{t_{\text{A}}}\sum_{t_{\text{B}}}\rho^{1/2}|s_{\text{A}}, s_{\text{B}}\rangle|\tilde{s}_{\text{A}}, \tilde{s}_{\text{B}}\rangle\langle t_{\text{A}}, t_{\text{B}}|\langle \tilde{t}_{\text{A}}, \tilde{t}_{\text{B}}|(\rho^{1/2})^{\dagger}.\label{eqA1}
\end{align}  
As shown in Eq.(\ref{eq12}), the partial trace of the variables corresponding to the spin B gives the renormalized extended density matrix $\hat{\rho}_{\text{A}}$ as follows:
\begin{align}
\hat{\rho}_{\text{A}}&\equiv {\text{Tr}}_{\text{B}}\hat{\rho}\equiv \sum_{\gamma_{\text{B}},\tilde{\gamma}'_{\text{B}}}\langle \gamma_{\text{B}},\tilde{\gamma}'_{\text{B}}|\hat{\rho}|\gamma_{\text{B}},\tilde{\gamma}'_{\text{B}}\rangle \notag
\\
&=\sum_{\gamma_{\text{B}},\tilde{\gamma}'_{\text{B}}}\sum_{s_{\text{A}}}\sum_{s_{\text{B}}}\sum_{t_{\text{A}}}\sum_{t_{\text{B}}}\langle \gamma_{\text{B}},\tilde{\gamma}'_{\text{B}}|\rho^{1/2}|s_{\text{A}}, s_{\text{B}}\rangle|\tilde{s}_{\text{A}}, \tilde{s}_{\text{B}}\rangle\langle t_{\text{A}}, t_{\text{B}}|\langle \tilde{t}_{\text{A}}, \tilde{t}_{\text{B}}|(\rho^{1/2})^{\dagger}|\gamma_{\text{B}},\tilde{\gamma}'_{\text{B}}\rangle\notag
\\
&=\sum_{\gamma_{\text{B}},\gamma'_{\text{B}}}\sum_{s_{\text{A}},t_{\text{A}}}\sum_{s_{\text{B}},t_{\text{B}}} \langle \gamma_{\text{B}}|\rho^{1/2}|s_{\text{B}}\rangle |s_{\text{A}},\tilde{s}_{\text{A}}\rangle\langle t_{\text{A}},\tilde{t}_{\text{A}}|\langle t_{\text{B}}|(\rho^{1/2})^{\dagger}|\gamma_{\text{B}}\rangle\delta_{\gamma'_{\text{B}},s_{\text{B}}}\delta_{\gamma'_{\text{B}},t_{\text{B}}}\notag
\\
&=\sum_{\gamma_{\text{B}},\gamma'_{\text{B}}}\sum_{s_{\text{A}},t_{\text{A}}}\langle \gamma_{\text{B}}|\rho^{1/2}|\gamma'_{\text{B}}\rangle |s_{\text{A}},\tilde{s}_{\text{A}}\rangle\langle t_{\text{A}},\tilde{t}_{\text{A}}|\langle \gamma'_{\text{B}}|(\rho^{1/2})^{\dagger}|\gamma_{\text{B}}\rangle.\label{eqA2}
\end{align}
Note that the density matrix $\rho^{1/2}$ does not operate to the tilde states. 
When the density matrix $\rho^{1/2}$ is expressed using the matrix elements $\{a_{\alpha_{\text{A}},\beta_{\text{B}},\alpha'_{\text{A}},\beta'_{\text{B}}}\}$ as shown in Eq.(\ref{eq13}), the term $\langle \gamma_{\text{B}}|\rho^{1/2}|\gamma'_{\text{B}}\rangle$ in Eq.(\ref{eqA2}) is obtained as follows:
\begin{align}
\langle \gamma_{\text{B}}|\rho^{1/2}|\gamma'_{\text{B}}\rangle &= \sum_{\alpha_{\text{A}},\beta_{\text{B}},\alpha'_{\text{A}},\beta'_{\text{B}}}a_{\alpha_{\text{A}},\beta_{\text{B}},\alpha'_{\text{A}},\beta'_{\text{B}}}|\alpha_{\text{A}}\rangle\langle \alpha'_{\text{A}}|\langle \gamma_{\text{B}}|\beta_{\text{B}}\rangle\langle\beta'_{\text{B}}|\gamma'_{\text{B}}\rangle\notag
\\
&= \sum_{\alpha_{\text{A}},\beta_{\text{B}},\alpha'_{\text{A}},\beta'_{\text{B}}}a_{\alpha_{\text{A}},\beta_{\text{B}},\alpha'_{\text{A}},\beta'_{\text{B}}}|\alpha_{\text{A}}\rangle\langle \alpha'_{\text{A}}|\delta_{\gamma_{\text{B}},\beta_{\text{B}}}\delta_{\beta'_{\text{B}},\gamma'_{\text{B}}}\notag
\\
&=\sum_{\alpha_{\text{A}},\alpha'_{\text{A}}}a_{\alpha_{\text{A}},\gamma_{\text{B}},\alpha'_{\text{A}},\gamma'_{\text{B}}}|\alpha_{\text{A}}\rangle \langle \alpha'_{\text{A}}|\label{eqA3}
\end{align}
Thus, inserting Eq.(\ref{eqA3}) to Eq.(\ref{eqA2}), we obtain the extended density matrix $\hat{\rho}_{\text{A}}$ as follows:
\begin{align}
\hat{\rho}_{\text{A}}&=\sum_{s_{\text{A}},t_{\text{A}}}\sum_{\gamma_{\text{B}},\gamma'_{\text{B}}}\sum_{\alpha_{\text{A}},\alpha'_{\text{A}}}\sum_{\beta_{\text{A}},\beta'_{\text{A}}}( a_{\alpha_{\text{A}},\gamma_{\text{B}},\alpha'_{\text{A}},\gamma'_{\text{B}}}|\alpha_{\text{A}}\rangle\langle \alpha'_{\text{A}}| ) |s_{\text{A}}\rangle|\tilde{s}_{\text{A}}\rangle\langle\tilde{t}_{\text{A}}|\langle t_{\text{A}}| ( |\beta'_{\text{A}}\rangle\langle \beta_{\text{A}}| a^{*}_{\beta_{\text{A}},\gamma_{\text{B}},\beta'_{\text{A}},\gamma'_{\text{B}}} )\notag
\\
&=\sum_{s_{\text{A}},t_{\text{A}}}\sum_{\gamma_{\text{B}},\gamma'_{\text{B}}}\sum_{\alpha_{\text{A}},\alpha'_{\text{A}}}\sum_{\beta_{\text{A}},\beta'_{\text{A}}}a_{\alpha_{\text{A}},\gamma_{\text{B}},\alpha'_{\text{A}},\gamma'_{\text{B}}} a^{*}_{\beta_{\text{A}},\gamma_{\text{B}},\beta'_{\text{A}},\gamma'_{\text{B}}} |\alpha_{\text{A}}\rangle |\tilde{s}_{\text{A}}\rangle\langle\tilde{t}_{\text{A}}|\langle \beta_{\text{A}}|\langle \alpha'_{\text{A}}|s_{\text{A}}\rangle\langle t_{\text{A}}|\beta'_{\text{A}}\rangle\notag
\\
&=\sum_{\alpha_{\text{A}},\alpha'_{\text{A}}}\sum_{\beta_{\text{A}},\beta'_{\text{A}}}\sum_{\gamma_{\text{B}},\gamma'_{\text{B}}}\sum_{s_{\text{A}},t_{\text{A}}}a_{\alpha_{\text{A}},\gamma_{\text{B}},\alpha'_{\text{A}},\gamma'_{\text{B}}} a^{*}_{\beta_{\text{A}},\gamma_{\text{B}},\beta'_{\text{A}},\gamma'_{\text{B}}} |\alpha_{\text{A}}\rangle|\tilde{s}_{\text{A}}\rangle\langle\tilde{t}_{\text{A}}|\langle \beta_{\text{A}}|\delta_{\alpha'_{\text{A}},s_{\text{A}}}\delta_{t_{\text{A}},\beta'_{\text{A}}}\notag
\\
&=\sum_{\alpha_{\text{A}},\alpha'_{\text{A}}}\sum_{\beta_{\text{A}},\beta'_{\text{A}}}\sum_{\gamma_{\text{B}},\gamma'_{\text{B}}}a_{\alpha_{\text{A}},\gamma_{\text{B}},\alpha'_{\text{A}},\gamma'_{\text{B}}} a^{*}_{\beta_{\text{A}},\gamma_{\text{B}},\beta'_{\text{A}},\gamma'_{\text{B}}} |\alpha_{\text{A}},\tilde{\alpha}'_{\text{A}}\rangle\langle\beta_{\text{A}},\tilde{\beta}'_{\text{A}}|\notag
\\
&=\sum_{\alpha_{\text{A}},\alpha'_{\text{A}}}\sum_{\beta_{\text{A}},\beta'_{\text{A}}}\left( \sum_{\gamma_{\text{B}},\gamma'_{\text{B}}}a_{\alpha_{\text{A}},\gamma_{\text{B}},\alpha'_{\text{A}},\gamma'_{\text{B}}} a^{*}_{\beta_{\text{A}},\gamma_{\text{B}},\beta'_{\text{A}},\gamma'_{\text{B}}} \right)|\alpha_{\text{A}},\tilde{\alpha}'_{\text{A}}\rangle\langle\beta_{\text{A}},\tilde{\beta}'_{\text{A}}|\notag
\\
&\equiv \sum_{\alpha_{\text{A}},\beta_{\text{A}},\alpha'_{\text{A}},\beta'_{\text{A}}}
b_{\alpha_{\text{A}},\beta_{\text{A}},\alpha'_{\text{A}},\beta'_{\text{A}}}
\left(|\alpha_{\text{A}}\rangle\langle\alpha'_{\text{A}}|\right)\left(|\tilde{\beta}_{\text{A}}\rangle\langle\tilde{\beta}'_{\text{A}}|\right).\label{eqA4}
\end{align}
This is nothing but the formulation (\ref{eq14}).
Of course, the elements $b_{\alpha_{\text{A}},\beta_{\text{A}},\alpha'_{\text{A}},\beta'_{\text{A}}}
$ are defind in Eq.(\ref{eq15}).
As shown in the above derivation, it is interesting to note that the matrix elements of $\hat{\rho}_{\text{A}}$, namely $b_{\alpha_{\text{A}},\beta_{\text{A}},\alpha'_{\text{A}},\beta'_{\text{A}}}$, include the fluctuation of the subsystem B through the summation $\sum_{\gamma_{\text{B}},\gamma'_{\text{B}}}$.
Once we obtain the matrix elements of the ordinary density matrix, we can calculate the matrix elements of $\hat{\rho}_{\text{A}}$ in the double Hilbert space using this formulation.
This is because our formulation requires, at most, the same computational time to diagonalize the ordinary density matrix, although the extended density matrix is defined in the double Hilbert space.

\bibliographystyle{model1a-num-names}

\begin{thebibliography}{00}
\bibitem{1}
U. Fano, Rev. Mod. Phys. A {\bf 42} (1957), 74.
\bibitem{2}
I. Prigogine et al., Chemica Scripta {\bf 4} (1973), 5.
\bibitem{3}
Y. Takahashi and H. Umezawa, Collect Phenom. {\bf 2} (1975), 55.
\bibitem{4}
M. Suzuki, J. Phys. Soc. Jpn. {\bf 54} (1985), 4483.
\bibitem{5}
M.Suzuki, ``Statistical Mechanics"iIwanami,2000,Tokyoj in Japanese.
\bibitem{6}
M. Suzuki, J. Stat. Phys. {\bf 42} (1986), 1047.
\bibitem{7}
A. E. Feiguin and S. R. White, Phys. Rev. B {\bf 72} (2005), 220401.
\bibitem{8}
W. Israel, Phys. Lett. {\bf 57A} (1976), 107.
\bibitem{9}
P. Benioff, J. Stat. Phys. {\bf 22} (1980), 563.
\bibitem{10}
P. Benioff, Phys. Rev. Lett. {\bf 48} (1982), 1581.
\bibitem{11}
W. K. Wooters and W. H. Zurek, Nature {\bf 299} (1996), 802.
\bibitem{12}
J. M. Maldacena, Phys. Rev. D {\bf 55} (1997), 7645; ibid Adv. Theor. Math. Phys. {\bf 2} (1998), 231; ibid Int. J. Theor. Phys. {\bf 38} (1999) 1113.
\bibitem{13}
S. Ryu and T. Takayanagi, Phys. Rev. Lett. {\bf 96} (2006), 181602.
\bibitem{14}
P. Calabrese and J. Cardy, J. Stat. Mech. (2004), P06002.
\bibitem{15}
A. Kitaev and J. Preskill, Phys. Rev. Lett. {\bf 96} (2006), 110404.
\bibitem{16}
M. Cadoni and M. Melis, Entropy {\bf 12} (2010), 2244.
\bibitem{17}
H. H. Lin, Commun. Theor. Phys. {\bf 55} (2011), 349.
\bibitem{18}
J. Ren, S. Zhu and X. Hao, J. Phys. B {\bf 42} (2009), 015504.
\bibitem{19}
G. Vidal, J. I. Latorre, E. Rico and A. Kitaev, Phys. Rev. Lett. {\bf 90} (2003), 227902.
\bibitem{20}
F. Verstraete, M. Popp and J. I. Cirac, Phys. Rev. Lett. {\bf 92} (2004), 027901.
\bibitem{21}
J. A. Hoyos, A. P. Vieira, N. Laflorencie and E. Miranda, Phys. Rev. B {\bf 76} (2007), 174425.
\bibitem{22}
F. Alet, S. Capponi, N. Laflorencie and M. Mambrini, Phys. Rev. Lett. {\bf 99} (2007), 117204.
\bibitem{23}
G. Refael and J. E. Moore, Phys. Rev. Lett. {\bf 93} (2004), 260602.
\bibitem{24}
H. Matsueda, Phys. Rev. E {\bf 85} (2012), 031101.
\bibitem{25}
M. Suzuki, Prog. Theor. Phys. {\bf 56} (1976), 1454.
\bibitem{26}
M. Suzuki, in: L. Accardi, W. Freundberg and M. Ohya (Eds), PQ-QP, QBIC, World Scientific, Singapore, 2008.
\bibitem{27}
M. Suzuki, Physica A {\bf 390} (2011) 1904, ibid {\bf 391} (2012) 1074. 
\end{thebibliography}

\end{document}